\documentclass[aps,prl,twocolumn,a4paper,10pt,notitlepage,footinbib,superscriptaddress,longbibliography]{revtex4-1}
\usepackage[english]{babel}
\usepackage{lmodern}
\usepackage[latin9]{inputenc}
\usepackage{endnotes}
\usepackage{amssymb,amsmath,amsfonts}
\usepackage{textcomp}
\usepackage{braket}
\usepackage{ulem}
\usepackage{soul}

\usepackage{graphicx}% Include figure files
\usepackage{dcolumn}% Align table columns on decimal point
\usepackage{bm}% bold math
\usepackage{xcolor}
\usepackage{soul}
\begin{document}

\title{Shapes and dynamic regimes of a polar active fluid droplet under confinement}

\author{A. Tiribocchi$^*$} 
\affiliation{Istituto per le Applicazioni del Calcolo CNR, via dei Taurini 19, 00185 Rome, Italy\\ Corresponding author: adrianotiribocchi@gmail.com}
\author{M. Durve}
\affiliation{Center for Life Nano Science@La Sapienza, Istituto Italiano di Tecnologia, 00161 Roma, Italy}
\author{M. Lauricella}
\affiliation{Istituto per le Applicazioni del Calcolo CNR, via dei Taurini 19, 00185 Rome, Italy}
\author{A. Montessori}
\affiliation{Department of Engineering, Universit\`{a} degli Studi Roma Tre, Via Vito Volterra 62, 00146 Rome, Italy}
\author{D. Marenduzzo}
\affiliation{Scottish Universities Physics Alliance, School of Physics and Astronomy, University of Edinburgh, Edinburgh EH9 3JZ, United Kingdom}
\author{S. Succi}
\affiliation{Istituto per le Applicazioni del Calcolo CNR, via dei Taurini 19, 00185 Rome, Italy}
\affiliation{Center for Life Nano Science@La Sapienza, Istituto Italiano di Tecnologia, 00161 Roma, Italy}
\affiliation{Department of Physics, Harvard University, Cambridge, MA, 02138, USA}

\begin{abstract}
Active droplets are artificial microswimmers built from a liquid dispersion by microfluidic tools and showing self-propelled motion. These systems hold particular interest for mimicking biological phenomena, such as some aspects of cell locomotion and collective behaviors of bacterial colonies, as well as for the design of droplet-based biologically inspired materials, such as engineered tissues.  Growing evidence suggests that geometrical confinement crucially affects their morphology and motility, but the driving physical mechanisms are still poorly understood. Here we study the effect of activity on a droplet containing a contractile polar fluid confined within microfluidic channels of various sizes. We find a surprising wealth of shapes and dynamic regimes, whose mechanics is regulated by a subtle interplay between contractile stress, droplet elasticity and microchannel width. They range from worm-like and cell-like shaped droplets displaying an oscillating behavior within wider channels to bullet-shaped droplets exhibiting rectilinear motion in narrower slits. Our findings support the view that geometrical confinement can provide a viable strategy to control and predict the propulsion direction of active droplets. It would be of interest to look for analogues of these motility modes in biological cells or in 
synthetic active matter.
\end{abstract}

\maketitle

\section{Introduction}

Active matter comprises systems in which the constituents building-blocks are capable of continuously converting chemical energy into mechanical work \cite{marchetti,ramaswamy}. An important paradigm of active matter are fluids of self propelled rod-like units with the tendency to assemble, giving rise to liquid crystals with nematic or polar order \cite{marchetti2}. Instances of these hierarchical soft structures (often termed active gels \cite{joanny}) are ubiquitous in nature and range from bacterial or algae suspensions \cite{arci,cates_rev} to mixtures of cytoskeletal filaments (such as actin or microtubules) and molecular motors \cite{sriram}. The interaction of these internal entities with each other and with the surrounding medium produces mechanical stress causing a surprising number of non-equilibrium effects, such as collective motion at large scales \cite{julicher,sumino,surrey}, spontaneous flows \cite{dogic,kruse}, turbulence at low Reynolds number \cite{wensink,alert} and exotic rheological responses \cite{fielding,clement,saintillan}. This instability ultimately depends upon the nature of the forces generated by the active particles which are treated, at the lowest order, as active force dipoles. If these forces are directed towards the center of mass of the particles, the material is contractile, otherwise it is extensile \cite{marchetti}.

Recent microfluidic advances have shown that the self-locomotion of micrometer-sized fluid droplets can be enabled by encapsulating an active gel within an emulsified dispersion of oil in water. Successful experimental realizations include, for example, stabilized bundles of microtubules cross-linked by kinesin (an extensile gel) adsorbed onto the interface of a droplet \cite{dogic,sagues3} or included at its interior \cite{sagues2}, and acto-myosin fibres (contractile gel) assembled within cell-sized droplets \cite{maeda}. Besides their biological importance as model tools to study complex biological processes, such as intracellular movements \cite{marchetti,giomi}, cell swimming \cite{tjhung1,yoshinaga} and crawling \cite{tjhung2,aranson3}, they may be ideal candidates for the design of bio-inspired active materials potentially useful in material science, for the assemble of engineered tissues from cell layers \cite{yeom_nat,giomi2}, and in pharmaceutics as microreactors for cargo delivery, including drug delivery and water remediation \cite{menglin2018,maass2021,gao1,gao2}.   

Under these circumstances, these self-motile droplets often navigate within confined microenvironments (such as porous viscoelastic tissues and capillary vessels), a situation that poses additional constraints, with respect to  unconfined systems, to their morphology and locomotion regimes, especially in the presence of narrow interstices \cite{int_bio,yamada}. Indeed, the presence of confining interfaces (or walls) can decisively affect the swimming motion due to unavoidable disturbances of the hydrodynamic flow field which could ultimately impact on droplet shape and stability. Such physiological-like conditions can be  partially reproduced in {\it in-vitro} experiments by means of suitable microfluidic devices in which an active droplet migrates within micrometer-sized channels  in a controlled manner \cite{maass,hokmabad,deblois2019,maeda,stamat2020,deblois2021,maass2021}. While  a large body of studies have been focused on investigating the physics of weakly confined swimmers  \cite{lauga,zottl,liu,acemoglu,fadda_epje,kanso}, much less is known in highly confined geometries despite their practical relevance.

In this paper, we numerically study the dynamics of a two-dimensional polar active fluid droplet navigating within  microchannels of various heights whose design draws inspiration from real microfluidic devices \cite{deblois2019,deblois2021}. Extensive lattice Boltzmann simulations show a striking variety of shapes and dynamic regimes whose formation relies upon a delicate balance between contractile stress, droplet elasticity (including those of liquid crystal and fluid interface) and degree of confinement (defined as the ratio between height of the channel and droplet diameter). While, under weak confinement, the droplet is generally found to exhibit an oscillatory self-propelled motion with shapes ranging from a rounded cell-like structure at small active doping to a stretched worm-like configuration at large activity, increasing the confinement considerably impacts on shapes and droplet trajectories. At intermediate confinement, the droplet either acquires a bullet-like profile with negligible oscillations at small activity or attains, once again, an oscillating cell-like shape displaying a unidirectional motion at higher active stress. Further increasing the confinement prevents the droplet motion, except for cases in which a high active stress fosters recurrent and periodic shape changes without producing any net displacement. The dynamical behaviour we find is richer than the one observed for confined squirmers~\cite{li2014,lintuvuori2016}, which can also entail sustained oscillations, because of the key role played by both shape deformations and internal fluid flow in our droplets, which are absent in squirmers. The motility regimes found here are characterized in terms of the structure of the flow field and of its coupling with the orientation of the active material and the droplet interface, as well as through elastic free energy changes associated with droplet interface and liquid crystal distortions. All these findings are finally summarised in a phase diagram providing an overview of the dynamics of a contractile fluid droplet swimming in microfluidic channels.

The paper is structured as follows. In the next section we shortly describe the hydrodynamic gel theory alongside some numerical aspects concerning the simulation setup. Afterwards we illustrate the numerical results, initially focusing on the dynamics of the active droplet in an unbounded medium and then  on that within microchannels of various heights. A selection of  shapes and motility regimes are discussed in three different confinement conditions, together with quantitative measurements of center of mass position and speed. The aforementioned phase diagram and some final remarks close the paper.

\section{METHODS}\label{methods}
\subsection{Free energy}
The physics of a polar active fluid droplet immersed in a passive Newtonian fluid can be described in terms of the following set of coarse-grained fields: 
(i) $\phi({\bf r},t)$, the concentration of the active material; (ii) ${\bf P}({\bf r},t)$ the polarization of the liquid crystal; (iii) $\rho({\bf r},t)$ and (iv) ${\bf u}({\bf r},t)$, respectively mass density and velocity field of the fluid. Our active droplet is designed as a circular region comprising a contractile gel ($\phi>0$, ${\bf P}\neq 0$), which can be experimental realized, for example, with bundles of actin filaments cross-linked with myosin motor  proteins \cite{marchetti,sagues}. This suspension is surrounded by a passive wet solvent ($\phi=0$ and ${\bf P}=0$) confined between two parallel flat walls. 

The equilibrium properties of the passive limit of this suspension are encoded in the following free energy density \cite{marenduzzo,tiribocchi2}
\begin{eqnarray}\label{free}
  f=&&\frac{a}{4\phi_{cr}^4}\phi^2(\phi-\phi_0)^2+\frac{k}{2}(\nabla\phi)^2\nonumber\\&&
  -\frac{\alpha}{2}\frac{(\phi-\phi_{cr})}{\phi_{cr}}|{\bf P}|^2+\frac{\alpha}{4}|{\bf P}|^4+\frac{\kappa}{2}(\nabla{\bf P})^2,
\end{eqnarray}
where the first two terms describe bulk and interfacial properties of a binary fluid mixture, while the remaining ones capture those of the polar liquid crystal. More specifically, the first term of Eq.\ref{free} guarantees the separation of the binary fluid into two components, $\phi = \phi_0$ (with $\phi_0$ approximately equal to $2$) inside the droplet and $\phi=0$ outside. The second term controls the interfacial tension of the drop, which is given by $\sigma=\sqrt{8ak/9}$ where $k$ and $a$ are positive constants. The terms multiplied by the constant $\alpha$ represent the bulk contributions associated with the liquid crystal phase and $\phi_{cr}=\phi_0/2$ is the critical concentration controlling the transition from the isotropic phase (outside the active drop) to the polar one (within the drop). Finally, the term in gradients of ${\bf P}$ accounts for the local deformations of the liquid crystal within the standard approximation of the single elastic constant $\kappa$ \cite{degennes}.

To summarize, at equilibrium we have the following phases minimizing the free energy $F=\int_V fdV$: a polar active phase located within the drop (where $\phi=\phi_0$ and ${\bf P}={\bf P}_{eq}$)
and a passive isotropic one (where $\phi=0$ and ${\bf P}=0$) outside; across the interface of the droplet, $\phi$ and ${\bf P}$ vary smoothly from $\phi=\phi_0$ and ${\bf P}={\bf P}_{eq}$ to $\phi=0$ and ${\bf P}=0$.

\subsection{Equations of motion}

The fluid velocity ${\bf u}$ is governed by the Navier-Stokes equations which, in the incompressible limit, read
\begin{equation}\label{cont_eq}
\nabla\cdot{\bf u}=0,
\end{equation}
\begin{equation}\label{nav_stok}
  \rho\left(\frac{\partial}{\partial t}+{\bf u}\cdot\nabla\right){\bf u}=-\nabla p + \nabla\cdot(\underline{\underline\sigma}^{active}+\underline{\underline\sigma}^{passive}).
\end{equation}
Here  $\rho$ is the density of the fluid and $p$ is the isotropic pressure. On the right hand side, $\underline{\underline\sigma}^{active}+\underline{\underline\sigma}^{passive}$ is the total stress tensor, given by the sum of active and passive terms. The former is \cite{ramaswamy}
\begin{equation}\label{act_st}
\sigma_{\alpha\beta}^{active}=-\zeta\phi\left(P_{\alpha}P_{\beta}-\frac{1}{d}|{\bf P}|^2\delta_{\alpha\beta}\right),    
\end{equation}
where Greek indexes denote Cartesian components, $d$ is the dimension of the system and  $\zeta$ is the activity, which is positive for extensile particles and negative for contractile ones \cite{marchetti}. In our simulations it is kept negative, meaning that the active gel contracts along the direction of the inner units. 

The passive stress comprises three further contributions, namely a viscous term $\sigma_{\alpha\beta}^{viscous}=\eta(\partial_{\alpha}v_{\beta}+\partial_{\beta}v_{\alpha})$ where $\eta$ is the shear viscosity, an elastic stress $\underline{\underline{\sigma}}^{elastic}$ due to bulk distortions of the liquid crystal and a surface tension term $\underline{\underline{\sigma}}^{interface}$.
The elastic contribution is
\begin{equation}\label{el_st}
\sigma_{\alpha\beta}^{elastic}=\frac{1}{2}(P_{\alpha}h_{\beta}-P_{\beta}h_{\alpha})-\frac{\xi}{2}(P_{\alpha}h_{\beta}+P_{\beta}h_{\alpha})-\kappa\partial_{\alpha}P_{\gamma}\partial_{\beta}P_{\gamma},
\end{equation}
where the constant $\xi$ controls the geometry of the active particles (positive for rod-like and negative disk-like ones) and it also determines the response of the liquid crystal under shear, which is either flow aligning (if $|\xi|>1$) or flow tumbling (if $|\xi|<1$).  As in previous works \cite{tjhung1,tjhung2}, we set $\xi>1$. The last term of Eq.\ref{el_st} represents the contribution of the Ericksen stress in the single elastic constant approximation \cite{degennes,teren}.

Finally, the interfacial stress is
\begin{equation}\label{int_st}
\sigma_{\alpha\beta}^{interface}=\left[\left(f-\phi\frac{\delta{\cal F}}{\delta\phi}\right)\delta_{\alpha\beta}-\frac{\partial f}{\partial(\partial_{\beta}\phi)}\partial_{\alpha}\phi\right].    
\end{equation}

The dynamics of the order parameters $\phi$ obeys a Cahn-Hilliard equation \cite{cahn,bray}
\begin{equation}\label{cahn_hilliard}
\frac{\partial\phi}{\partial t}+\nabla\cdot(\phi{\bf u})=M\nabla^2\mu,
\end{equation}
where $M$ is the mobility -- chosen to be constant for simplicity -- and $\mu=\delta  F/\delta\phi$ is the chemical potential.

Finally, the evolution equation for the polarization ${\bf P}({\bf r},t)$ is given by \cite{marchetti}
\begin{equation}\label{P_eq}
\frac{\partial {\bf P}}{\partial t} + ({\bf u}\cdot\nabla){\bf P}=-\underline{\underline\Omega}\cdot{\bf P}+\xi\underline{\underline D}\cdot {\bf P}-\frac{1}{\Gamma}\frac{\delta F}{\delta {\bf P}},   
\end{equation}
where $\underline{\underline{D}}=(\underline{\underline W}+\underline{\underline W}^T)/2$ and $\underline{\underline\Omega}=(\underline{\underline W}-\underline{\underline W}^T)/2$ are the symmetric and antisymmetric part of the velocity gradient tensor $W_{\alpha\beta}=\partial_{\beta}u_{\alpha}$. The last term, multiplied by the rotational viscosity $\Gamma$, is the molecular field ${\bf h}=\delta F/\delta {\bf P}$ which governs the relaxation dynamics of the liquid crystal. 

\subsection{Numerical implementation and mapping to physical units}

Following previous studies \cite{tjhung1,tiribocchi3}, Eqs.(\ref{cont_eq})-(\ref{nav_stok})-(\ref{cahn_hilliard})-(\ref{P_eq}) are solved via a hybrid lattice Boltzmann approach \cite{succi,tiribocchi3}, in which 
a standard LB method is used for Eqs. (\ref{cont_eq}) and (\ref{nav_stok}) and a finite difference scheme for Eqs. (\ref{cahn_hilliard}) and (\ref{P_eq}). In particular, 
finite difference operators are computed using a stencil representation, an approach guaranteeing the necessary isotropy with the underlying two dimensional lattice \cite{tiribocchi_pre,thampi}. 
The simulations are performed on two-dimensional rectangular meshes, in which the horizontal direction is kept constant at $L_y=500$ lattice sites, while the vertical one $L_z$ is varied between $45$ and $135$ lattice sites. Periodic boundary conditions are set along the $y$-axis and two parallel flat walls are positioned at $z=0$ and $z=L_z$. Here we impose no wetting for $\phi$ (i.e. $\phi|_{z=0,L_z}=0$) and  no-slip conditions for ${\bf u}$ (i.e. $u|_{z=0,L_z}=0$), while no specific anchoring is set for ${\bf P}$ at the walls.
In this microfluidic channel we place a contractile droplet initialized as a circular region of radius $R$ where $\phi({\bf r},0)\simeq 2$ and ${\bf P}({\bf r},0)={\bf P}_y({\bf r},0)$ (with $|{\bf P}_y|=1$) for $r\le R$, while $\phi({\bf r},0)\simeq 0$ and ${\bf P}({\bf r},0)=0$ outside, for $r>R$.  
In our runs we have kept $R$ constant and equal to $45$ lattice sites, thus the confinement parameter $\lambda=L_z/2R$ ranges from $0.5$ (which we refer to as high confinement) to $1.5$ (which we refer to as low confinement).

If not stated otherwise, the thermodynamic parameters have been fixed to the following values: $a=0.04$, $k=0.06$, $\alpha=0.1$, $\kappa=0.04$, $M=0.1$, $\xi=1.1$, $\Gamma=1$, $\eta\simeq 1.67$. The first two values, in particular, determine the surface tension $\sigma=\sqrt{8ak/9}\simeq 0.045$ and the interface width $\delta=2\sqrt{2k/a}\simeq 3.5$. Note that, for simplicity, the viscosity of the background fluid and the one of the fluid inside the drop are equal. This approximation could be, in principle, relaxed by letting $\eta$ depend on $\phi$ \cite{tjhung2}. Also, lattice spacing and integration timestep are  set to $\Delta x=1$ and $\Delta t=1$. Finally, the contractile activity $|\zeta|$ has been varied approximately between $5\times 10^{-4}$ and $2.5\times 10^{-3}$: these values generally ensure a spontaneous motion while minimizing the chance of droplet breakups, events more likely to occur in the presence of intense flows. 

Simulation parameters can be mapped onto real physical units by fixing  length, time and force scales to the following values: $L=1\mu$m,  $T=10$ms and $F=100$nN (in simulation units these scales are equal to one) \cite{tjhung1,tjhung2}.  Thus our system would approximately simulate a microfluidic channel of length  $\sim 0.5$mm in which the polar active fluid droplet of diameter $D\simeq 90\mu$m has shear viscosity $\eta\simeq 1.5$kPa$\cdot$s, while the liquid crystal has effective elastic constant $\kappa\simeq 4$nN and rotational viscosity $\Gamma\simeq 1$kPa$\cdot$s. Also, the  diffusion coefficient is $D_{\phi}=Ma\simeq 0.4$ $\mu$m$^2$/s. Finally, the speed $v$ of the droplet in our simulation varies between $\sim 10^{-4}$ to $\sim 10^{-3}$, values roughly corresponding to $0.1-1$ $\mu m/s$. These numbers ensure that the Reynolds number $Re=\rho v D/\eta$ remains equal to or lower than $0.1$ and the Capillary number $Ca=v\eta/\sigma$ is below $0.1$.

\section{Results}
We start off by describing the typical dynamics observed in an unbounded system and then we present a selection of shapes and dynamic regimes observed in the weak ($\lambda>1$), intermediate ($\lambda\simeq 1$) and strong ($\lambda<1$) confinement for various $\zeta$ focusing, in particular, on the fluid-structure interaction. We finally provide a full phase diagram of a contractile fluid droplet moving within microchannels of various heights. 

\subsection{Contractile droplet in an unbounded passive fluid}

\begin{figure*}[htbp]
\includegraphics[width=1.0\linewidth]{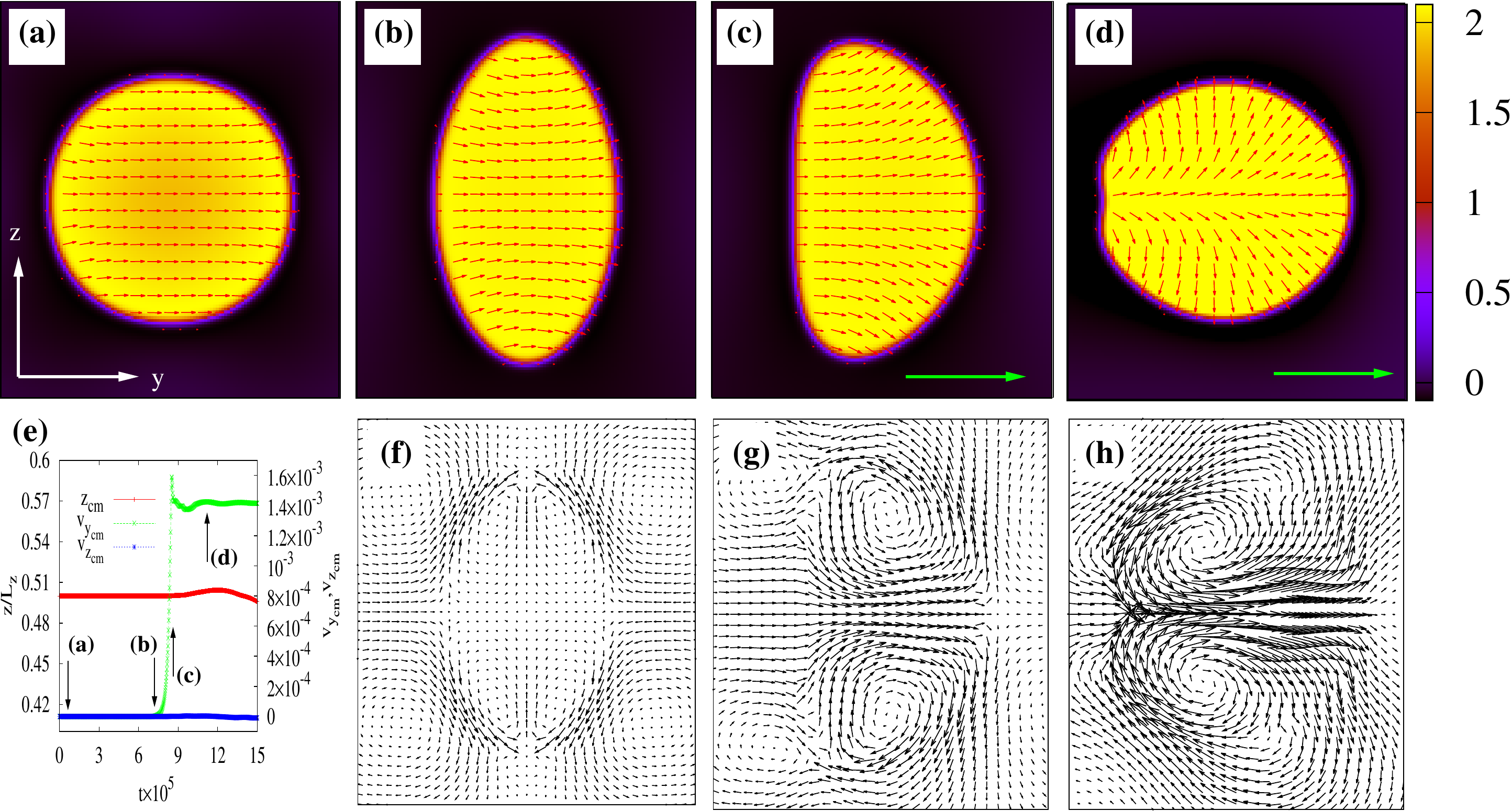}
\caption{{\bf Autonomous motion of a polar active fluid droplet in an unbounded  passive Newtonian fluid for $\zeta=-8\times 10^{-4}$}. (a) Initial configuration of an equilibrated passive droplet and of the polarization, which is almost everywhere parallel to the $y$ direction. (b)-(f) Once the activity is turned on, the droplet temporarily elongates perpendicularly to ${\bf P}$ due to the contractile activity and its associated four-vortex fluid flow pattern, which pulls the fluid along the equator and expels it  longitudinally. (c)-(g) The uniform polarization becomes unstable with respect to splay deformations. This instability is due to the large contractile stress which favours the spontaneous formation of two counter-rotating vortices triggering the motion of the droplet. (d)-(h) The droplets acquires a stable rounded shape and moves unidirectionally sustained by the two internal vortices. The color bar ranges from $\phi\simeq 0$ (black) to $\phi\simeq 2$ (yellow), red arrows represent the direction of the polarization ${\bf P}$ and the green arrow indicates the direction of motion. Black arrows show the local direction of the velocity field ${\bf u}$. This applies to all figures. (e) Time evolution of the $z$ component of the center of mass (left axis) and of $y$ and $z$ components of its speed (right axis). Multimedia view.}
\label{fig1N}
\end{figure*}

The mechanism leading to the autonomous motion of a contractile fluid droplet has been extensively studied, for example, in Refs.\cite{tjhung1,hawkins}. In Fig.\ref{fig1N} (Multimedia view M1) we shortly outline the physics in unbounded systems and in a regime in which the motion is unidirectional. The droplet is initialized as a passive circular region in which ${\bf P}$ is uniform and aligned along the $y$ direction (Fig.\ref{fig1N}a). Then $\zeta$ is turned on and is set to a value triggering spontaneous motion. However, before this occurs, the contractile stress (whose strength is controlled by $\zeta$) sets a four-roll mill flow which stretches the droplet longitudinally without causing any net motion (Fig.\ref{fig1N}b,f). This state results from the balance  between interfacial tension and liquid crystal elasticity which hinder shape deformations, and contractility favouring hydrodynamic instability. Afterwards, the contractile stress overcomes the resistance to deformations mediated by liquid crystal and fluid interface, thus  destabilizing the polarization pattern and yielding an outward splay distortion associated with the formation of two counter-rotating vortices in the flow field, which trigger droplet motion (Fig.\ref{fig1N}c,g). Concomitantly, the droplet acquires a temporary crescent shape along the direction of motion. At late times, the droplet shape becomes rounded (Fig.\ref{fig1N}d) and the motion proceeds at constant speed parallel to the $y$ direction (Fig.\ref{fig1N}e), sustained by the two fluid vortices (Fig.\ref{fig1N}h).

\begin{figure*}[htbp]
\includegraphics[width=1.0\linewidth]{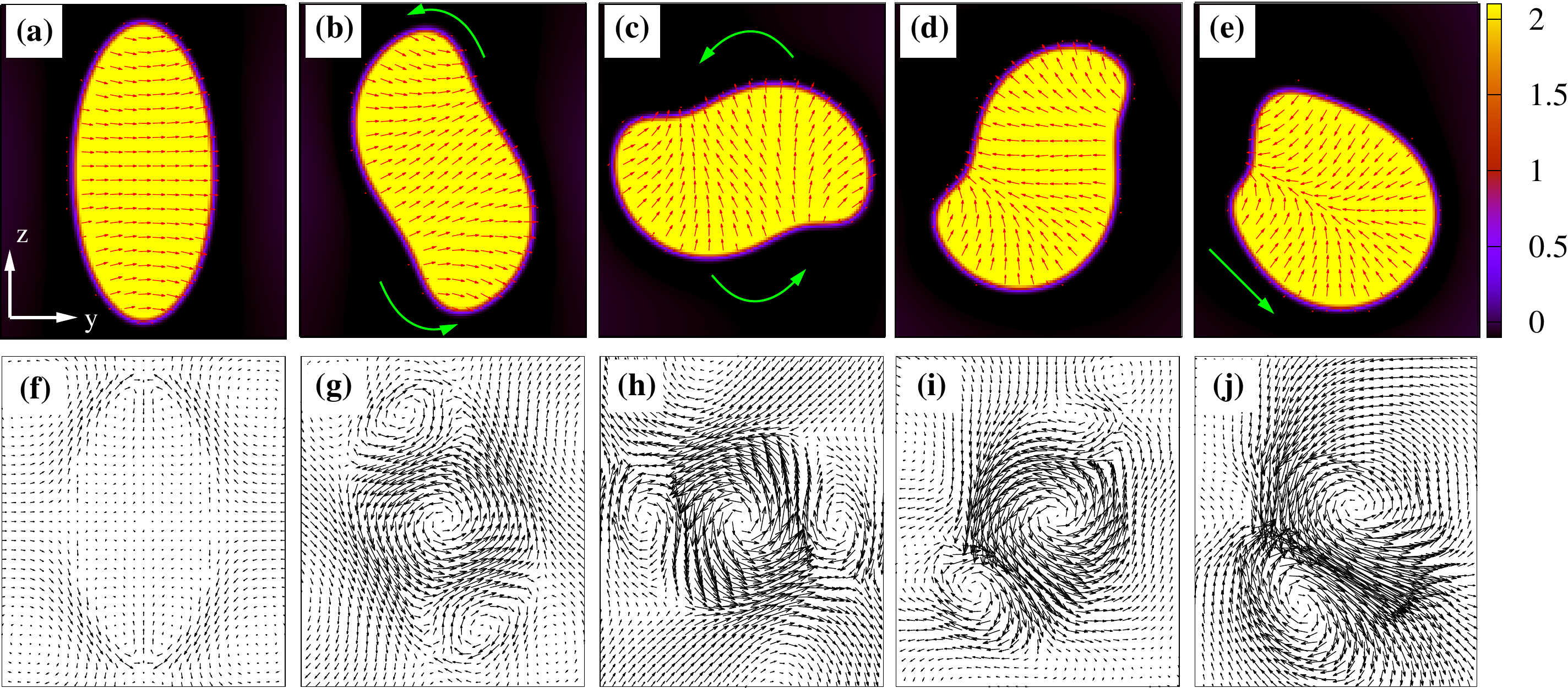}
\caption{{\bf Autonomous motion of a polar active fluid droplet in an unbounded passive Newtonian fluid for $\zeta=-10^{-3}$}. Once the activity is turned on, the fluid flow stretches the droplet longitudinally (a-f). Then, the contractile stress destabilizes the polarization, which remains uniform in the bulk of the droplet and splays at the extremities (b-g). Three contiguous fluid vortices, a large one rotating counterclockwise and two smaller ones rotating clockwise, emerge and trigger the rotation of the droplet (c-h). Afterwards, the large vortex in the middle and the small one on the top merge into a large double-vortex structure leading to self propulsion, while the polarization progressively rearranges to yield a single splay deformation pointing inwards (d-i and e-j). Multimedia view.}
\label{fig2N}
\end{figure*}

Increasing $|\zeta|$ generally leads to higher order instabilities in which one observes, unlike the previous case, the concurrent presence of regions where neighboring vectors of ${\bf P}$ splay outwards ($\nabla\cdot{\bf P}>0$) and inwards ($\nabla\cdot{\bf P}<0$). In Fig.\ref{fig2N} (Multimedia view M2) we show, for example, the onset of spontaneous motion of an active droplet with $\zeta=-10^{-3}$. Once again, the droplet initially stretches longitudinally due to the four-vortex fluid flow (Fig.\ref{fig2N}a,f) and is subsequently destabilized by the contractile stress. Here the polarization shows three different patterns: It remains basically uniform in the middle of the droplet, where a large counterclockwise vortex emerges, and splays inwards and outwards at the extremities, where two small clockwise vortices appear (Fig.\ref{fig2N}b,g). This is in agreement with the findings discussed in Ref.\cite{fialho} where it is shown that the dynamics of an active droplet is decisively affected by the anchoring of the liquid crystal at the fluid interface: a normal orientation leads to a single splay deformation (as in Fig.\ref{fig1N}d) while a tangential one stabilizes a couple of splay distortions (as in Fig.\ref{fig2N}b). Our drop contains a hybrid of these two conditions: the orientation of the polar field is perpendicular almost everywhere except at the extremities of the drop, where it is preferentially tangential (see Fig.\ref{fig2N}a). Thus, if the contractile stress is sufficiently high, at the top and bottom of the drop two splay deformations, associated with two non-collinear and opposing active forces, emerge and produce a torque which leads to rotation (Fig.\ref{fig2N}c,h). However, such motion lasts for a short period of time since the perpendicular orientation prevails and leads, once again, to a single splay deformation (Fig.\ref{fig2N}d,e). Note that now the splay points inwards following the direction of the polar field. At the onset of the instability, this choice (whether inwards or outwards) is made at random, depending on the breaking of the polar inversion symmetry \cite{tjhung1}. Once the rotation is over, the vortex in the middle gradually merges with a small one into a large double-vortex structure (Fig.\ref{fig2N}h,i), finally yielding a unidirectional motion (Fig.\ref{fig2N}e,j).

These results suggest that, although at the onset of motion some dynamic features (shape, structure of polarization and velocity field) depend on $\zeta$, in the steady state the droplet is essentially found to move along a rectilinear path without relevant perturbations, for a wide range of values of active parameter (see also the movie SM1 for $\zeta=-1.5\times 10^{-3}$ in the supplementary material). In the next section we show that, under confinement, this picture is dramatically altered and the active droplet displays a multifarious range of shapes and dynamic regimes which are absent in unbounded systems.

\subsection{Weak confinement}

\begin{figure*}[htbp]
\includegraphics[width=1.0\linewidth]{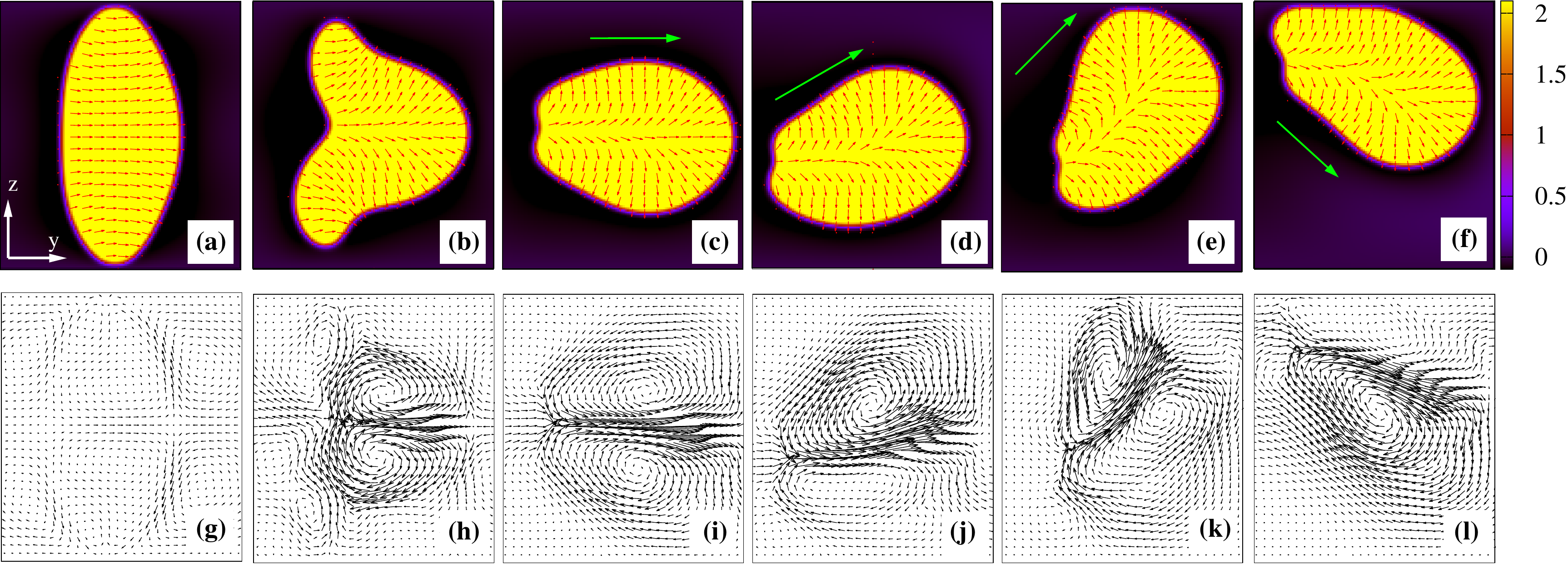}
\caption{{\bf Shapes and dynamic behavior of a polar active fluid droplet with $\zeta=-1.5\times 10^{-3}$ and $\lambda\simeq 1.5$}. Once $\zeta$ is turned on, the droplet elongates (a) due to the usual four-vortex structure of the fluid in its surrounding (g). Afterwards, the contractile stress destabilizes this configuration, favouring the formation of two roughly symmetric branches in which the polar field gives rise to splay deformations, pointing outwards at the front of the droplet and inwards at its rear (b). Alongside the two large counter-rotating vortices, two short-lived small ones emerge on the back (h). The droplet then acquires a slightly elongated cell-like shape (c) self propelled by the usual double-vortex velocity field (i). As the droplet moves downwards and approaches the wall (d), the vortex at the bottom weakens (j), an effect that reverses the motion towards the upper wall (e,k). Finally, the droplet hits the top wall and turns downwards, once again driven by a single fluid vortex surviving far from the boundary (f,l). This figure corresponds to cell-like shape and unidirectional oscillating motion of the phase diagram of Fig.\ref{fig3} (green/triangles). Multimedia view.}
\label{fig7}
\end{figure*}

In Fig.\ref{fig7} (and Multimedia view M3) we show the active droplet dynamics for $\lambda\simeq 1.5$ and $\zeta=-1.5\times 10^{-3}$. The onset of the instability resembles the one observed in unconfined systems: the droplet initially stretches longitudinally due to the four-roll mill flow only mildly affecting the orientation of the polarization field (Fig.\ref{fig7}a,g). Afterwards, the contractile stress deforms the droplet, which displays two temporary symmetric branches. Here the polarization splays outwards at the front, where a couple of counter-rotating fluid vortices emerges, and inwards at the rear, where two smaller vortices form (Fig.\ref{fig7}b,h). These ones then merge with the leading ones giving rise to two large fluid recirculations pushing the droplet forward (Fig.\ref{fig7}c,i). In this configuration, the droplet attains a slightly elongated shape and the polar field globally shows an outward splay distortion.

However, the presence of the walls  crucially alters the perfect symmetry of the vortices, an effect that modifies droplet shape and orientation of the polarization finally causing a deviation from the rectilinear trajectory. Indeed, a weak perturbation (caused by the interaction between fluid and walls) may drive, for example, the droplet close to the bottom wall (Fig.\ref{fig7}d,j) where the nearby fluid vortex lessens due to the momentum sink (no slip conditions are set the boundaries). On the contrary, the vortex far from the wall strengthens, allowing the droplet to turn towards the middle of the channel (where the double vortex is temporarily restored, Fig.\ref{fig7}e,k)). Once in close contact with the top wall, the droplet weakly flattens and turns downwards driven, once again, by a single clockwise vortex (Fig.\ref{fig7}f,l). During this process, its shape remains globally akin to that observed in the short rectilinear motion of Fig.\ref{fig7}c, while the polarization  shows a long-lasting splay deformation pointing outwards and parallel to the direction of motion. Such dynamics repeats itself, giving rise to a regular and persistent oscillatory behavior in which the droplet moves forward periodically hitting opposite walls (see also Fig.\ref{fig9} where position and speed of the center of mass are plotted for several values of $\zeta$).

\begin{figure*}[htbp]
\includegraphics[width=1.0\linewidth]{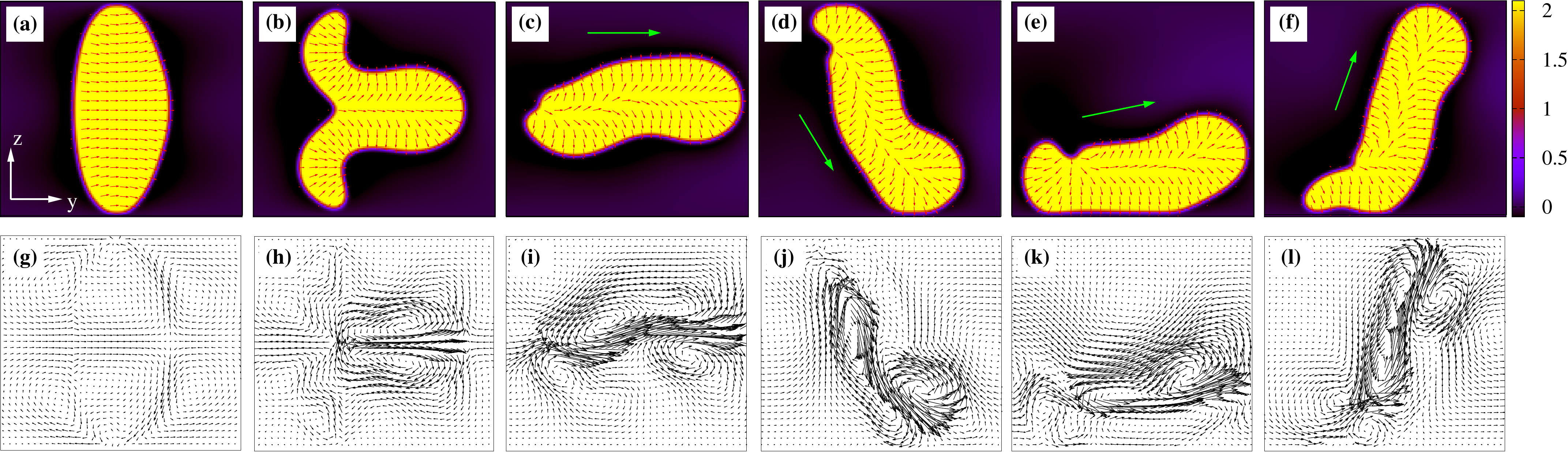}
\caption{{\bf Shapes and dynamic behavior of a polar active fluid droplet with $\zeta=-2.5\times 10^{-3}$ and $\lambda\simeq 1.5$}. The contractile stress destabilizes the droplet, which stretches longitudinally (with a uniform polarization) (a) and then lengthens horizontally acquiring a marked bifurcation at the rear (b) (with a global splay deformation of the polarization). Here the velocity field turns  to a stretched double vortex (h) from an initial four-roll pattern (g). Afterwards, the two branches merge and the droplet temporarily moves forward (c), propelled by a large single stream supported by smaller fluid vortices on both sides (i). The interaction of the fluid with the walls destabilizes the rectilinear motion favouring an oscillatory dynamics where a highly elongated worm-like droplet moves towards the bottom wall (d), remains close to it (e) and then turns upwards (f). Such motion is mainly guided by a stream crossing the droplet longitudinally, while side vortices of various size continuously break and reform (j,k,l). This figure corresponds to worm-like shape and unidirectional oscillating motion of the phase diagram of Fig.\ref{fig3} (orange/triangles). Multimedia view.}
\label{fig8}
\end{figure*}

\begin{figure*}[htbp]
\includegraphics[width=1.0\linewidth]{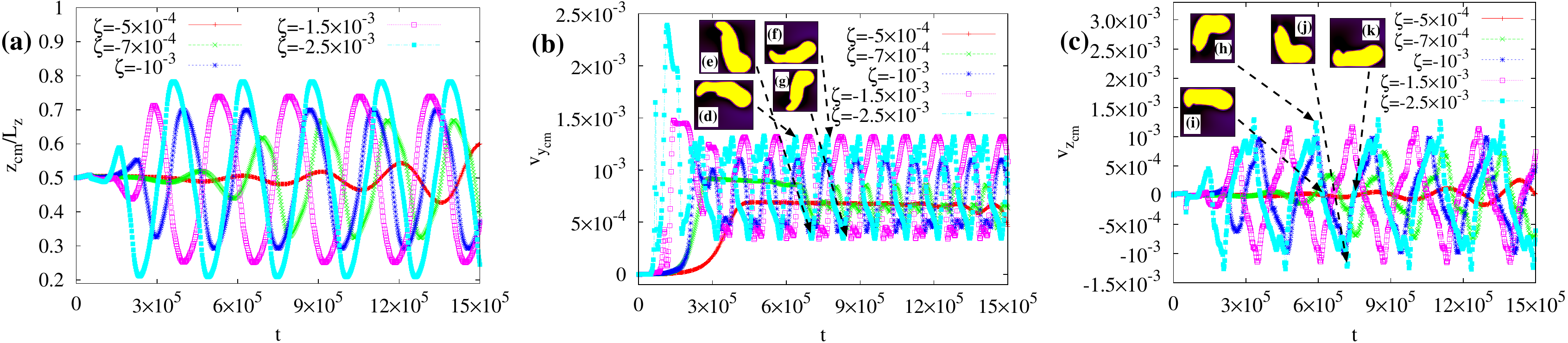}
\caption{{\bf Position and speed of the droplet center of mass under weak confinement}. (a) Time evolution of the $z$-component of the center of mass of the droplet for various values of activity. Increasing $|\zeta|$ augments both amplitude and frequency of oscillations. (b)-(c). Time evolution of $y$ and $z$ components of the speed of the center of mass. The snapshots represent instantaneous configurations of the droplet for $\zeta=-2.5\times 10^{-3}$ when $v_{y_{cm}}$ has a local maximum (d,f) and a local minimum (e,g), and when $v_{z_{cm}}$ has a local maximum (h), is approximately zero (i,k) and has a local minimum (j).}
\label{fig9}
\end{figure*}

For stronger contractile stresses, the droplet undergoes higher shape deformations albeit the essential features of the periodic motion are preserved. In Fig.\ref{fig8} (Multimedia view M4) we show the dynamics of the active droplet for $\lambda\simeq 1.5$ and $\zeta=-2.5\times 10^{-3}$. The onset of the instability occurs through steps akin to the previous case, i.e. a vertical elongation with uniform polarization caused by the four-vortex flow (Fig.\ref{fig8}a,g) and a subsequent double tail deformation with symmetric splay distortions of the polar field, an arrangement due to a couple of neighboring counter-rotating vortices placed at the front plus two separate smaller ones located at the rear (Fig.\ref{fig8}b,h). Afterwards,  the two tails merge leading to a wormlike shape, whose propulsion is sustained by a combination of an intense stream crossing the drop midsection plus highly stretched fluid vortices located on both sides (Fig.\ref{fig8}c,i). Once again, the complex interaction between  active medium and surrounding fluid with the microchannel destabilizes the rectilinear motion. Indeed, the droplet initially turns downwards (Fig.\ref{fig8}d,j) and then moves in close contact with the wall driven by a large unidirectional stream which is, in turn, flanked by stretched vortices continuously breaking and reforming (Fig.\ref{fig8}e,k).  As in the case of lower contractility, the fluid vortex near the bottom wall dramatically weakens, thus allowing the droplet to turn  and move towards the upper wall (Fig.\ref{fig8}f,l). During such process, the polar field essentially preserves a large splay deformation pointing outwards and parallel to the direction of motion.

These simulations show that, under confinement, increasing the active stress yields two crucial features: morphologically, it favours larger shape deformations becoming permanent at late times; dynamically, it fosters a persistent periodic motion characterized by regular oscillations.

Such behavior can be quantitatively captured by tracking position and speed of the center of mass of the droplet, as shown in Fig.\ref{fig9}. While at low values of $|\zeta|$ they display only mild oscillations, at higher values their amplitude and frequency increase, an effect caused by the stronger flows produced by the higher contractility. Note, in particular, that $v_{y_{cm}}$ attains a local maximum when the droplet moves near the walls (top or bottom) and turns (downwards or upwards), while a local minimum is found when the droplet hits the wall almost perpendicularly (Fig.\ref{fig9}d,e,f,g). On the contrary, $v_{z_{cm}}$ achieves the largest values slightly after the collision against the wall, while it rapidly diminishes and becomes almost negligible when the drop remains in close contact with the walls (Fig.\ref{fig9}h,i,j,k).

\subsection{Intermediate confinement}
\begin{figure*}[htbp]
\includegraphics[width=1.0\linewidth]{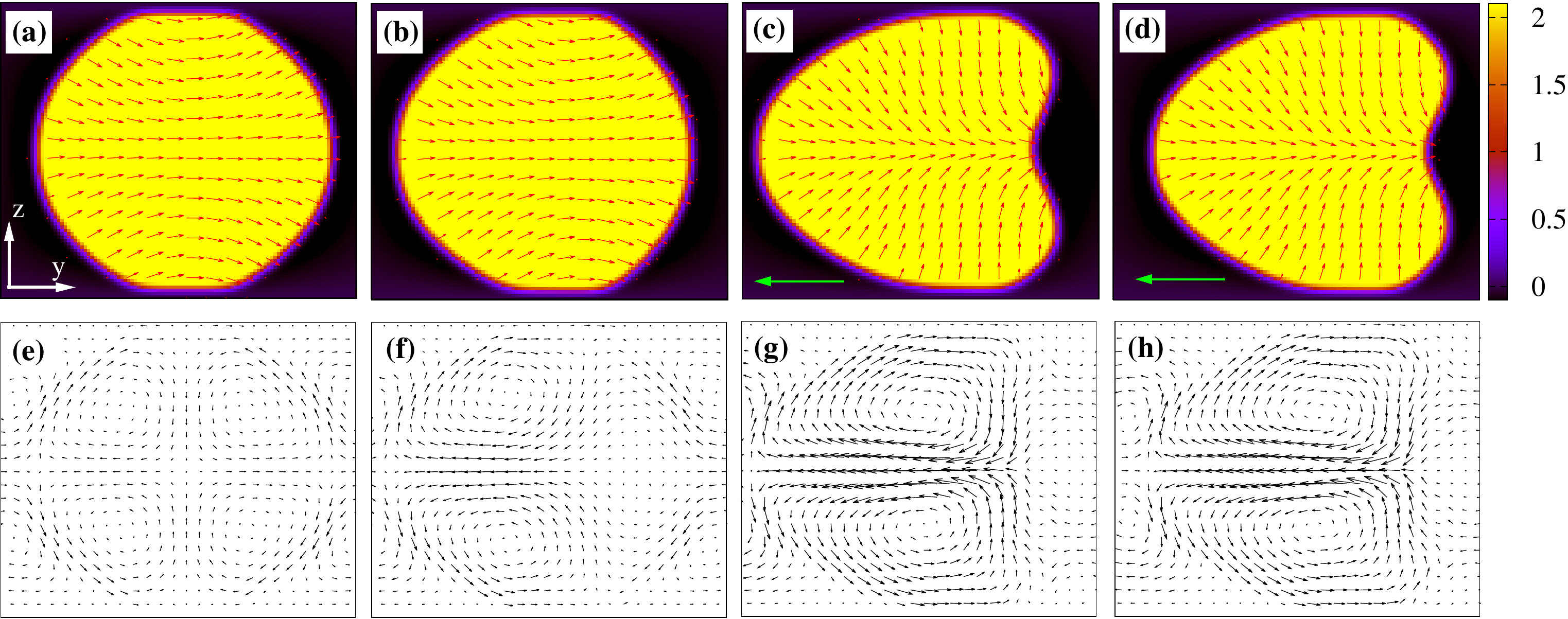}
\caption{{\bf Shapes and dynamic behavior of a polar active fluid droplet with $\zeta=-1.25\times 10^{-3}$ and $\lambda\simeq 1$}. The contractile stress destabilizes the squeezed droplet (a,b), which acquires motion along the $y$-direction attaining a projectile shape at the steady state (c,d). The four-roll fluid velocity, observed for the non-motile droplet, gradually turns into the typical double vortex structure (f,g,h), which propels the droplet forward. Note that, in the steady state, the polarization splays inwards rather than outwards. This figure corresponds to bullet shape and unidirectional motion of the phase diagram of Fig.\ref{fig3} (blue/asterisks). Multimedia view.}
\label{fig11}
\end{figure*}

\begin{figure*}[htbp]
\includegraphics[width=1.0\linewidth]{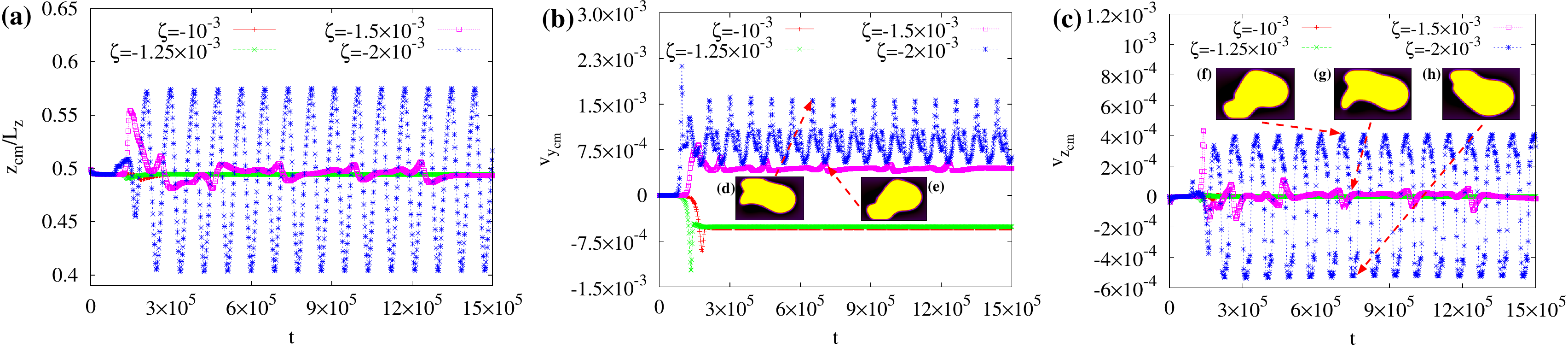}
\caption{{\bf Position and speed of the droplet center of mass under intermediate confinement.} (a) Time evolution of the $z$-component of the center of mass of the active droplet for various values of $\zeta$ and $\lambda\simeq 1$.  (b)-(c). Time evolution of $y$ and $z$ components of the speed of the center of mass. The snapshots represent instantaneous configurations of the droplet for $\zeta=-2\times 10^{-3}$ when $v_{y_{cm}}$ has local maximum (d) and a local minimum (e), and when $v_{z_{cm}}$ has a local maximum (f), is approximately zero (g) and has a local minimum (h).}
\label{fig12}
\end{figure*}

Increasing the confinement parameter $\lambda$ gives rise to further shapes and dynamic regimes. In Fig.\ref{fig11} (Multimedia view M6) we show an active droplet swimming within a microfluidic channel with $\lambda\simeq 1$ and $\zeta=-1.25\times 10^{-3}$. Under these conditions, the fluid interface of an initially static droplet slightly flattens near the walls due to the absence of wetting, allowing for the stabilization of a thin layer of background passive fluid (the black one). Once the activity is switched on, the contractile stress produces the usual four-vortex flow which, later on, turns into a double vortex pattern triggering an inward splay deformation. Unlike the previous regime, the higher confinement suppresses the shape oscillations and stabilizes  a bullet-like shaped droplet moving along a rectilinear trajectory (see Fig.\ref{fig12} for position and speed of center of mass). However, this dynamics holds as long as the active forces are  mild.

If the active stress increases, the droplet undergoes larger shape deformations due to the strengthening of the spontaneous flow, an effect that leads, once again, to a periodic motion akin to that observed under weak confinement. This is shown in the movie SM2 in the supplementary material, where an active fluid droplet (with $\zeta=-2\times 10^{-3}$) attains a cell-like shape moving, for a short period of time, along a rectilinear direction. Then, the interaction with the walls destabilizes this motion yielding an oscillatory dynamics in which the droplet moves forward periodically hitting against both walls. In this system $v_{y_{cm}}$ attains its maximum roughly in the middle of the channel (after the drop detaches from the wall) where $v_{z_{cm}}$ is approximately null, whereas the minimum values occur when the droplet bumps against the wall, where $v_{z_{cm}}$ has its largest (absolute) values (see Fig.\ref{fig12}). Note, finally, that the amplitude of the oscillations of $v_{z_{cm}}$ is considerably smaller than the one observed in the low confinement regime (for similar values of $\zeta$), essentially because the shorter height of the microchannel prevents the droplet from attaining a higher transversal speed. This is also in agreement with experimental results discussed in Ref.\cite{deblois2021}, where the behavior of self-propelled droplets confined in micro-capillary devices of different cross-sections is investigated.

In the next section we will show that, in contrast to dynamic behaviors discussed so far, in a strong confinement regime the active droplet either displays an oscillating stationary dynamics without an appreciable displacement or stays motionless.  

\subsection{Strong confinement}

\begin{figure*}[htbp]
\includegraphics[width=1.0\linewidth]{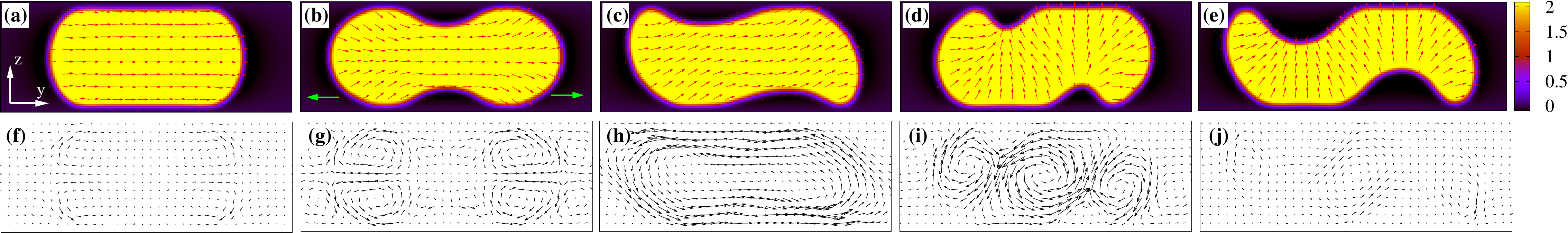}
\caption{{\bf Shapes and dynamic behavior of a polar active fluid droplet with $\zeta=-2.25\times 10^{-3}$ and $\lambda\simeq 0.6$}. The contractile stress destabilizes a highly confined droplet which stretches and acquires a temporary dumbbell-like configuration, made of two protuberances propelled along opposite directions by two separate couples of counter-rotating vortices (a,b,f,g). Afterwards, the droplet gradually acquires an S-like shape, sustained by a large recirculating flow pattern which later on splits into three smaller vortices (c,d,e,h,i,j). The polar field shows a splay deformation which, in the S-shaped drop, points both outwards and inwards in separate regions of the drop. This figure corresponds to S-like shape and oscillating stationary dynamics of the phase diagram of Fig.\ref{fig3} (black/circles). Multimedia view.}
\label{fig13}
\end{figure*}

As an example of dynamics under strong confinement, in Fig.\ref{fig13} (Multimedia view M8) we show the behavior obtained for $\zeta=-2.25\times 10^{-3}$ and $\lambda\simeq 0.6$. A highly squeezed droplet, initially placed in the middle of the microchannel, elongates producing two separate bulges which are pushed along opposite directions  by couples of counter-rotating fluid vortices. (Fig.\ref{fig13}a,b,f,g). These, in turn, favour the formation of splay distortions of the liquid crystal, pointing outwards at the front and inwards ad the rear. The fluid protuberances continuously merge and reform due to the competition between  contractile stress  (which tends to destabilize the drop) and the elasticity of droplet and liquid crystal (which oppose shape deformations), until the former overcomes the latter.  When this occurs, the droplet undergoes a profound reshaping in which it initially expands uniformly due to the merging of the vortices into a single large recirculating pattern (Fig.\ref{fig13}c,h), and then it gradually acquires a temporary S-like shape. Here  three smaller vortices emerge, one in the middle and two at the extremities of the drop (Fig.\ref{fig13}d,e,i,j). However, the contractile stress destabilizes this state finally leading to a steady-state oscillatory behavior between S-like shaped droplets mirroring each other. In these configurations, the liquid crystal exhibits the usual splay deformation, pointing both outwards and inwards in different regions of the droplet. 

If $|\zeta|$ decreases, shape oscillations are suppressed. This is shown in the movie SM3 in the supplementary material, obtained for $\zeta=-10^{-3}$ and $\lambda\simeq 0.7$. Here the contractile stress is not sufficient to trigger any relevant change of liquid crystal orientation and shape of the droplet, which undergo only  weak deformations due to internal feeble fluid recirculations. Besides the activity, the physics may be also affected by the fluid interface which, under strong confinement, would occupy a larger portion of channel. In our simulations we have $2\xi/L_z\simeq 0.15$, where $\xi\simeq 3.5$ and $L_z=45$. In the supplementary material we show that, keeping $\xi$ constant and $\lambda$ fixed to values pertaining to the high confinement regime, the results discussed in this section hold even if larger channels (thus lower values of $\xi/L_z$) are considered. It is finally worth noting that, unlike the dynamic behavior discussed so far, the motion through confined geometries could be potentially achieved through a tank-treading mode driven by active stress, with no need of traction on the walls and regardless of the slip conditions \cite{liverpool_soft}. This physics is generally different from that of our active droplet, in which the instability is generated by a spontaneous flow while no slip and no wetting are imposed at the walls.

\subsection{Shape deformations and variations of the liquid crystal orientation}

The results discussed so far point to a picture where the observed range of shapes and dynamic regimes  emerge from a subtle interplay between active stress, elasticity of the drop (including that of the liquid crystal and of the interface) and confining conditions.

\begin{figure*}[htbp]
\includegraphics[width=1.0\linewidth]{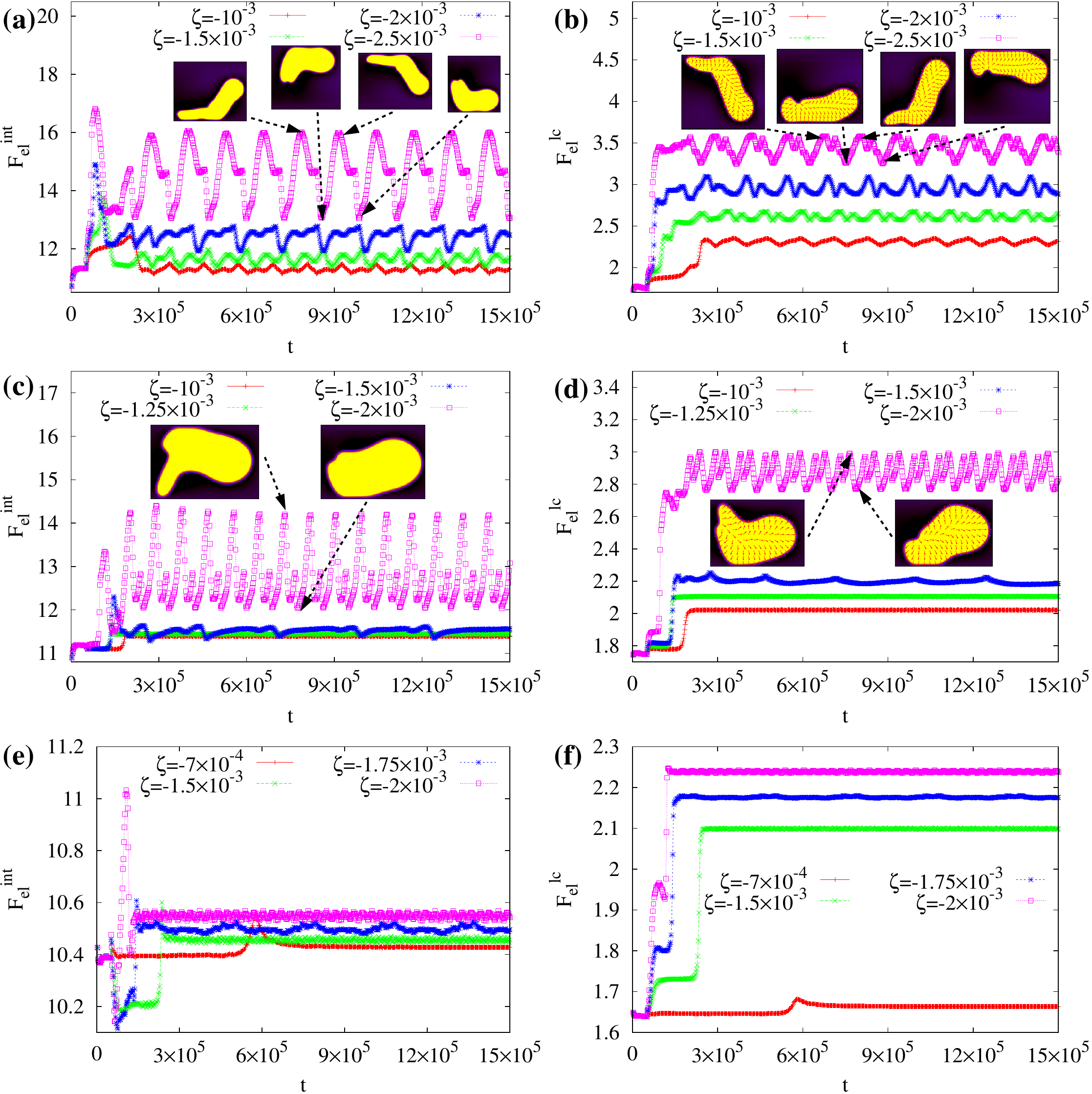}
\caption{{\bf Elastic free energy of the droplet}. Time evolution of  $F_{el}^{int}=\int_V k/2(\nabla\phi)^2 dV$ (left column) and $F_{el}^{lc}=\int_V \kappa/2(\nabla{\bf P})^2 dV$ (right column) for various  $\zeta$ and for $\lambda\simeq 1.5$ (a,b), $\lambda\simeq 1$ (c,d) and $\lambda\simeq 0.7$ (e,f). The insets show instantaneous configurations of $\phi$ and ${\bf P}$ at local maxima and minima for $\zeta=-2.5\times 10^{-3}$ (a,b) and $\zeta=-2\times 10^{-3}$ (c,d). While at low and intermediate values of $\lambda$ the free energies show long-lasting oscillations for a sufficiently large contractile stress, at high confinement these oscillations are essentially suppressed, regardless of the strength of the contractility.}
\label{fig6}
\end{figure*}

A more quantitative evaluation of morphological changes and variations of the liquid crystal orientation in these regimes can be obtained by monitoring the elastic free energy of the fluid interface $F_{el}^{int}=\int_V k/2(\nabla\phi)^2dV$ and that of the liquid crystal $F_{el}^{lc}=\int_V\kappa/2(\nabla{\bf P})^2dV$. Although these quantities only give a coarse-grained description of such modifications, their tracking over time provides a direct link with speed variations of the droplet and with the structure of the fluid velocity of the mixture. In Fig.\ref{fig6} we plot the time evolution of $F_{el}^{int}$ and $F_{el}^{lc}$ for $\lambda\simeq 1.5$ (top row), $\lambda\simeq 1$ (middle row), $\lambda\simeq 0.7$ (bottom row) and for different values of $\zeta$. If the contractile stress is augmented, both $F_{el}^{int}$ and $F_{el}^{lc}$ show a systematic increase, an indication that shape modifications and liquid crystal distortions become larger. More specifically, in a larger channel (top row)  high shape deformations and relevant liquid crystal distortions (maxima of Fig.\ref{fig6}a,b) occur when the droplet crosses the channel moving towards the opposite wall (with increasing transversal speed and decreasing longitudinal one, see for example Fig.\ref{fig9}b,c), whereas they lessen (minima of Fig.\ref{fig6}a,b) as the droplet retracts and moves coasting the walls (here the longitudinal speed increases while the transversal one is negligible). These results are essentially consistent with the fact that the spontaneous flow weakens near the wall, thus the droplet would naturally tend to acquire a more rounded shape and the liquid crystal to orient more uniformly. For intermediate values of $\lambda$ (Fig.\ref{fig6} middle row) the previous picture,  i.e. larger shape deformations and higher liquid crystal distortions in the bulk rather than near the walls, basically holds although such a distinction becomes slightly more elusive since droplet collisions against the walls become more frequent due to the reduced height of the microchannel.  

Note finally that, at low and intermediate values of $\lambda$, the free energies display recurring oscillations of increasing amplitude, a behavior overall akin to that observed for position and speed of the droplet. On the contrary, under strong confinement oscillations become negligible since the droplet either moves unidirectionally without appreciable shape changes at the steady state or stays motionless (unless stationary oscillations emerge, such as those described in Fig.\ref{fig13}). 

\subsection{Phase diagram}

\begin{figure*}[htbp]
\includegraphics[width=1.0\linewidth]{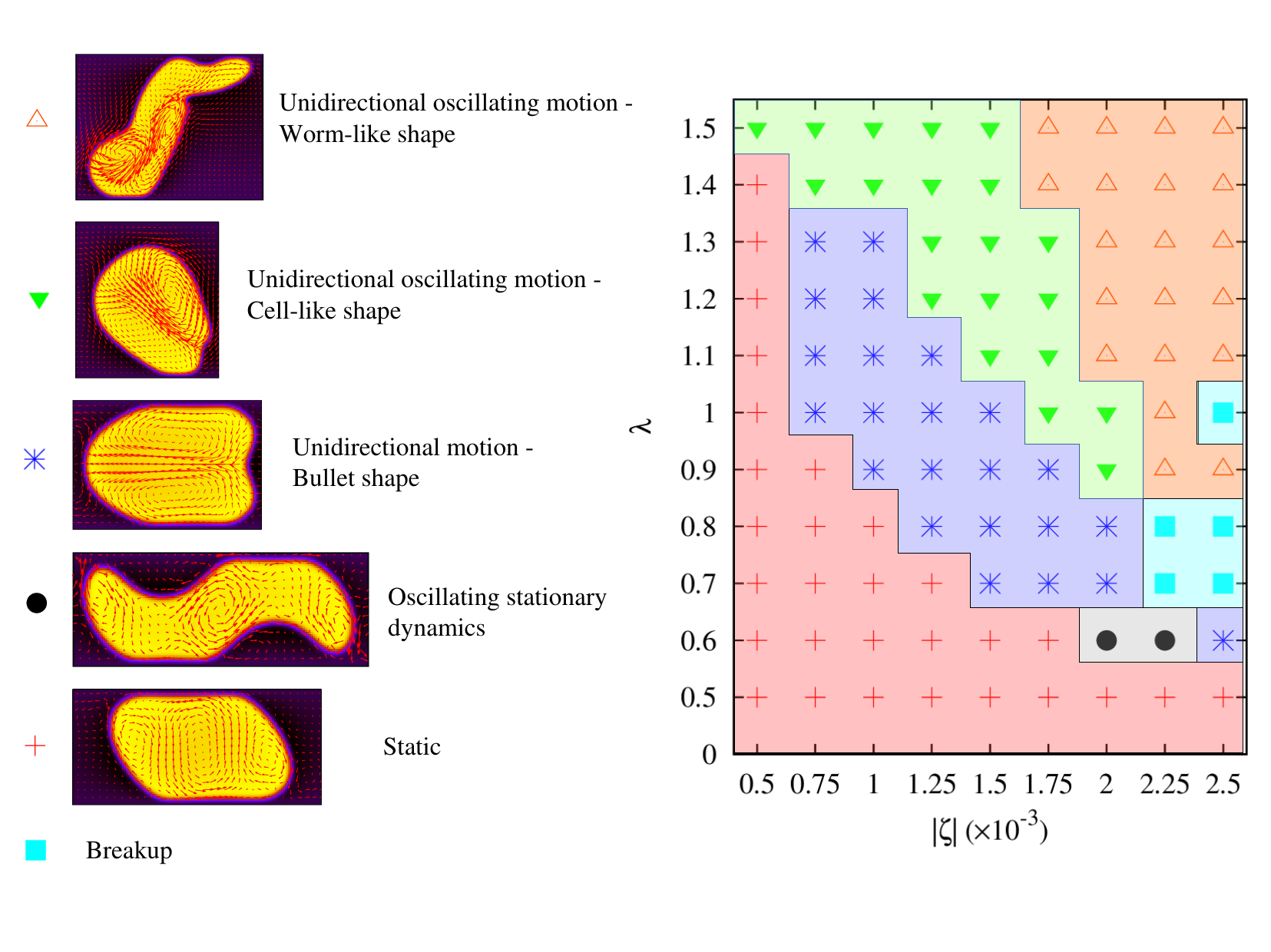}
\caption{{\bf Phase diagram of a polar active droplet for different values of confinement parameter $\lambda=L_z/D$ and activity $\zeta$.} Left: Selection of shapes of a contractile fluid droplet moving within  microchannels of various height. Red arrows show the velocity field within the droplet and in its surrounding. Right: Under strong confinement ($\lambda\lesssim 0.7$), for low values of $|\zeta|$ the droplet remains static (red/plus), while for high values of $|\zeta|$ it either displays stationary oscillations (black/circles) or acquires a bullet-like shape (blue/asterisk) moving unidirectionally. Also, occasional breakups (light blue/square) occur for very high values of $|\zeta|$. Under intermediate confinement ($0.8\lesssim\lambda\lesssim 1.2$), the droplet exhibits a larger variety of shapes and dynamic regimes, including cell-like (green/triangle) and worm-like (orange/triangle) structures for intermediate and high values of $|\zeta|$, respectively. In all such cases the droplet moves unidirectionally and displays a persistent oscillatory dynamics. At lower values of $|\zeta|$, the droplet shows, once again, a bullet shape moving without oscillations or remains motionless. Under weak confinement ($\lambda\gtrsim 1.3$), droplets self propel with long-lasting oscillations while their shapes mainly range from cell-like ones for lower $|\zeta|$ to stretched worm-like ones (orange/triangle) for higher $|\zeta|$.}
\label{fig3}
\end{figure*}

The complex dynamic scenario previously illustrated can be summarized in the phase diagram of Fig.\ref{fig3}, obtained by varying $|\zeta|$ and $\lambda$. 

For $\lambda\lesssim 0.7$ (strong confinement) and $|\zeta|<1.25\times 10^{-3}$ the highly squeezed droplet stays at rest, essentially because the contractile stress is too weak to deform the liquid crystal and the droplet interface in order to induce any net motion. On the contrary, for increasing  values of $|\zeta|$ the spontaneous flow becomes high enough to trigger a unidirectional motion of the droplet which acquires, at the steady state, a bullet-like shape, a structure closely resembling that of a passive fluid droplet subject to a pressure-driven flow~\cite{tiribocchi2}. Further increasing $|\zeta|$ may either determine the formation of long-lasting and periodic shape oscillations without motility or occasional breakups caused by intense fluid flows. 

For $0.8\lesssim\lambda\lesssim 1.2$ (intermediate confinement) the bullet shaped droplet moving along a rectilinear trajectory becomes the dominating regime at low/intermediate values of contractile activity, whereas the non-motile regime gradually shrinks towards very low values of $|\zeta|$. This occurs because, in larger channels, the 
constraining effect of the walls becomes progressively milder than that produced within narrower channels, thus morphological deformations and modifications of liquid crystal orientation can be more easily induced by an equally intense spontaneous flow. Alongside these dynamic regimes, two further ones emerges for increasing $|\zeta|$. For $1.5\times 10^{-3}\lesssim|\zeta|\lesssim 2\times 10^{-3}$, at late times the  droplet is found to display a unidirectional oscillating motion  with a shape resembling that of a stretched swimming cell. As discussed in the previous sections, such dynamics results from a complex fluid-structure interaction in which the presence of the walls alters the symmetric double-vortex structure of the fluid velocity, an effect that simultaneously fosters considerable changes of shape and orientation of the liquid crystal. This is particularly relevant for a high contractile stress, since the spontaneous flow is expected to impact deeper on the droplet dynamics. Indeed, at higher values of $|\zeta|$  the motion exhibits, once again, oscillatory features but the droplet attains a highly stretched worm-like shape moving at a larger speed.

Finally, for $\lambda\gtrsim 1.3$, the droplet is generally found to display a periodic dynamics for roughly all values of $|\zeta|$ explored in our simulations. However, while for $|\zeta|\gtrsim 1.5\times 10^{-3}$ the drop attains, at the steady state, the worm-like shape previously mentioned,  for smaller values of $|\zeta|$ a weaker flow field mitigate the deformations, yielding a shape now resembling that of a rounded cell (akin to the one observed in unconfined systems).

\section{Conclusions}
In conclusion, we have numerically studied the dynamics of a polar active fluid droplet migrating within microchannels of different heights, in which the setup mimics conditions reproducible in microfluidic experiments. Our droplet comprises a contractile gel, representing a fluid containing force dipoles with long-range orientational order, confined by interfacial tension and immersed in a passive Newtonian fluid. At the walls of the channel we set no-slip conditions for the fluid velocity and no wetting for the fluid interface. 

We have shown that, under confinement, the droplet displays a wealth of shapes and non-trivial dynamic behaviors resulting from a subtle interplay between active stress, elasticity of liquid crystal and droplet interface plus fluid-structure interactions. While the onset of motion is determined by a single mechanism (i.e. spontaneous flow produced by the active material plus splay deformations of the liquid crystal) regardless of confinement conditions, at the steady state five different dynamic regimes are observed under weak, intermediate and strong confinement, characterized by distinctive morphological and motility features. Under weak confinement the droplet exhibits a unidirectional oscillatory motion acquiring, at the steady state, either a cell-like shape (for low/intermediate active stress) or an elongated worm-like structure (for large active stress). Dialling up the degree of confinement, morphological changes generally diminish while the frequency of impacts against the walls considerably augments. In this regime the droplet shows either an oscillating-free rectilinear motion with a bullet-like profile or the periodic one with a cell-like shape previously mentioned. Under high confinement, motility is overall suppressed except for large contractile mixtures. On a general basis, a periodic behavior alongside with considerable shape deformations emerge when the degree of confinement remains intermediate/low and the contractility is intermediate/high, otherwise the droplet shows simpler dynamic features.  The formation of such striking variety of shapes and the associated motility regimes crucially depend upon the structure of the fluid velocity which, during the oscillatory motion, exhibits two main patterns: i) a couple of counter-rotating vortices if the droplet moves far from the walls  and ii) a single vortex with a large stream crossing the drop midsection if the motion occurs near the walls. The former is the typical structure observed in self-propelled droplets moving along a predefined direction while the latter allows the droplet to turn and detach from the walls. Under high confinement, where oscillations are generally negligible due to the lack of sufficient space, a net displacement would be solely triggered by the double fluid vortex. It is worth noting that boundary conditions can considerably alter this picture, especially when the environment is highly constrained. A different wettability, for example, would modify the equilibrated shape of a passive droplet, thus very likely affecting dynamics and morphology of the active one. In Ref.\cite{liverpool_soft,liverpool_prl} it has been also suggested that a tractionless tank-treading mechanism can lead to self-propulsion regardless of the slip at the walls. Understanding whether slip conditions, different from the ones used in this work, can favour droplet motion induced by a spontaneous flow needs further investigation. In addition, although our results seem to suggest that, at fixed activity, increasing the thickness of the channel favours droplet motion (overall in agreement with Ref.\cite{joanny_epl}), it would be of interest to evaluate to what extent this holds by changing boundary conditions.

Active droplets, like the ones described in this work, may be realized using water-in-oil emulsions including extensile or contractile materials, such as microtubule-motor suspensions adsorbed onto the drop interface and yielding an effectively 2D active gel (as in \cite{dogic}) or actomyosin filaments dispersed within the drop (as in \cite{maeda}). The constrained setup could be designed using a microfluidic channel with flat walls functionalized with suitable surface chemistry techniques \cite{stroka,munoz} to ensure appropriate boundary conditions. This system could partially reproduce the contraction-based motility of cells (such as human breast cancer cells) migranting through an extracellular matrix within microchannels \cite{lammer}. Indeed, these active droplets have been previously shown to capture some aspects of cell locomotion, such as swimming in an unbounded fluid \cite{tjhung1,tjhung3} and crawling on a solid subtrate \cite{tjhung2}, albeit they remain distant from a real cell \cite{aranson3}. Nonetheless, their simplified architecture could provide a robust platform for the manufacturing of biomimetic synthetic swimmers with the propensity to move. In this respect our results, besides being useful for a deeper understanding the physics of these objects, suggest that their direction of locomotion can be more easily controlled and predicted when moving within suitable confined geometries  (such as those whose design is inspired to physiological environments), a scenario particularly relevant in several technological applications, ranging from drug delivery to the design of soft composite materials. Finally, although some recent research has addressed the dynamics of 3D active droplets in unconfined systems \cite{tjhung1,ruske}, much less is known about the physics in a constrained geometry. Understanding the effect produced by confinement on these objects remains open and will be investigated in future works.

\section*{Data availability}
Data available on request from the authors.

\section*{Competing interests}
The authors declare no competing interests. 

\section*{Acknowledgments}
A. T., M. D., M. L., A. M. and S. S. acknowledge funding from the European Research Council under the European Union's Horizon 2020 Framework Programme (No. FP/2014-2020) ERC Grant Agreement No.739964 (COPMAT). 

\newpage
\section*{Supplementary Material}
The Supplementary Material contains a section discussing the effect of the interface thickness under high confinement and three movies. Movie SM1 shows the active droplet dynamics in an unbounded medium for $\zeta=-1.5\times 10^{-3}$, movie SM2 shows the active droplet dynamics for $\lambda\simeq 1$  and $\zeta=-2\times 10^{-3}$ and movie SM3 shows the active droplet dynamics for $\lambda\simeq 0.7$ and $\zeta=-10^{-3}$.

\subsection*{Effect of interface thickness in high confinement regime}
  In the main text we have shown that, under high confinement (i.e. $\lambda\lesssim 0.7$), the active droplet acquires motion along a rectilinear trajectory for large contractile mixtures, while it is generally static for lower values of $\zeta$ (see also Fig.10 of the main text). Besides activity, the physics in this regime could be also affected by the interface thickness, which would occupy a non-negligible portion of the microchannel.

\begin{figure}[htbp]
\includegraphics[width=1.0\linewidth]{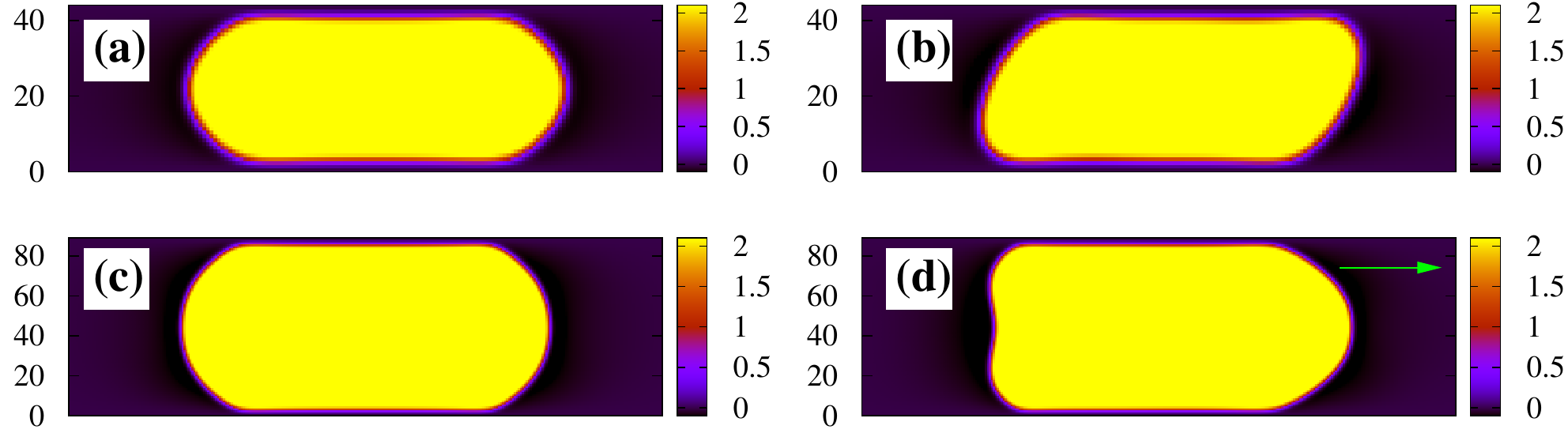}
\caption{Steady state configurations for (a) $\zeta = 10^{-3}$, $L_z = 45$, $D = 90$ (b) $\zeta = 2 \times 10^{-3}$, $L_z = 45$, $D = 90$, (c) $\zeta = 5 \times 10^{-4}$, $L_z = 90$, $D = 180$, (d) $\zeta = 7.5 \times 10^{-4}$, $L_z = 90$, $D = 180$. Here $D$ is the diameter of the corresponding circular droplet. In all cases $\lambda = L_z /D \simeq 0.5$ and $\xi \simeq 3.5$. The green arrow indicates the direction of motion.}
\label{figS1}
\end{figure}

In Fig.\ref{figS1} we show a sequence of steady state configurations where (a) $\zeta = 10^{-3}$ , $L_z = 45$, (b) $\zeta = 2 \times 10^{-3}$, $L_z = 45$, (c) $\zeta = 5 \times 10^{-4}$, $L_z = 90$ and (d) $\zeta = 7.5 \times 10^{-4}$, $L_z = 90$. In all cases $\lambda\simeq 0.5$, with $D = 90$ in  (a,b) and $D = 180$ in  (c,d). Also, the interface thickness has been kept fixed to $\xi\simeq  3.5$. In Fig.\ref{figS1}a,b one has $2\xi/L_z\simeq 0.15$ and, at the steady state, the drop is generally non-motile (for the range of values of $\zeta$ considered in the paper). These results are discussed in the phase diagram of Fig.10 of the main text.   In Fig.\ref{figS1}c,d one has $2\xi/L_z\simeq 0.078$ and, at the steady state, the droplet is either static (hence pertaining to the same class of Fig.\ref{figS1}a,b) for low values of $\zeta$ or motile (akin to the bullet shape) for higher values.  Note that, if $L_z$ increases, one needs larger droplets, containing more active material than smaller ones, to keep $\lambda$ constant; thus, $\zeta$ should concurrently decrease to reproduce conditions observed for smaller $L_z$.
On a general basis, the physics of these systems is controlled by a Cahn-like number $Ch = \xi/L_z$ and the Ericksen number $Er = \zeta L^2_z /\kappa$, where $\kappa$ is the elastic constant of the liquid crystal. If $Er > 1$, the active forces would be large enough to overcome the elasticity of the liquid crystal and potentially destabilize the droplet (although not necessarily inducing spontaneous motion, as in the case discussed here, but just a mild elongation). In the figure we have $Er\simeq 50$ (a), $Er\simeq 100$ (b), $Er\simeq 100$ (c) and $Er\simeq 150$ (d). Thus,  increasing $L_z$ requires a rescaling of $Er$ to capture the physics in similar regimes.

\subsection*{List of supplementary movies}
\begin{figure}[htbp]
\includegraphics[width=1.0\linewidth]{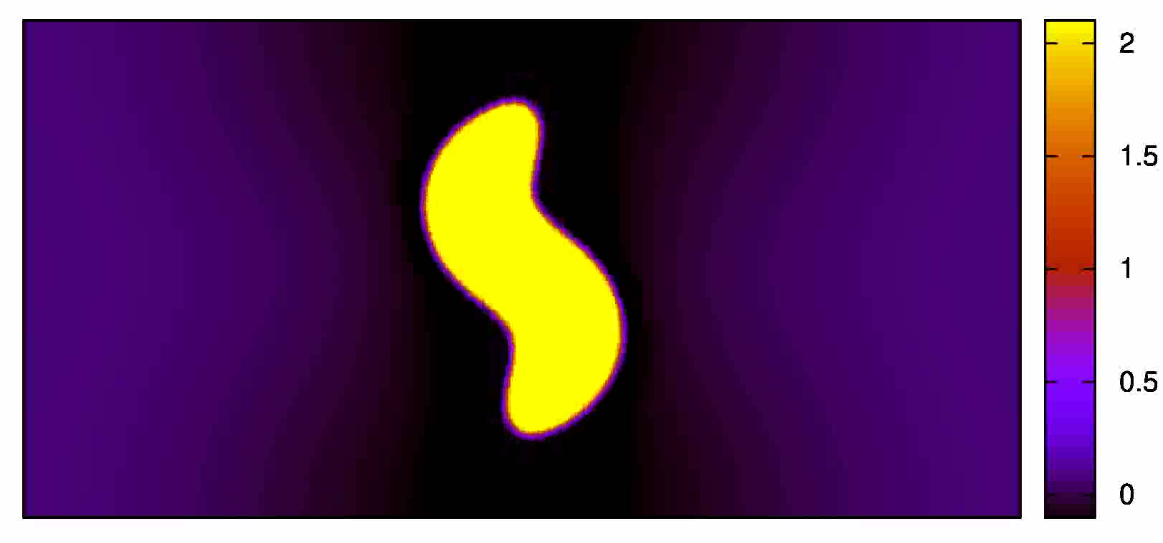}
\caption{Autonomous motion of a polar active fluid droplet in an unbounded passive Newtonian fluid for $\zeta=-1.5\times 10^{-3}$. The dynamics shares features similar to the ones discussed for $\zeta=-10^{-3}$ (Fig.2 of the main text).}
\label{figS2}
\end{figure}
\begin{figure}[htbp]
\includegraphics[width=1.0\linewidth]{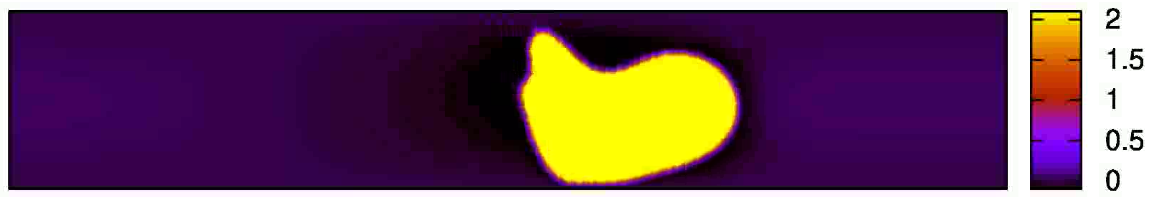}
\caption{Shapes and dynamic behavior of a polar active fluid droplet with $\lambda\simeq 1$ and $\zeta=-2\times 10^{-3}$. The motion is akin to that of Fig.3 of the main text. This state corresponds to a cell-like shape and unidirectional oscillating motion of the phase diagram of Fig.10 (green/triangles) of the main text.}
\label{figS3}
\end{figure}
\begin{figure}[htbp]
\includegraphics[width=1.0\linewidth]{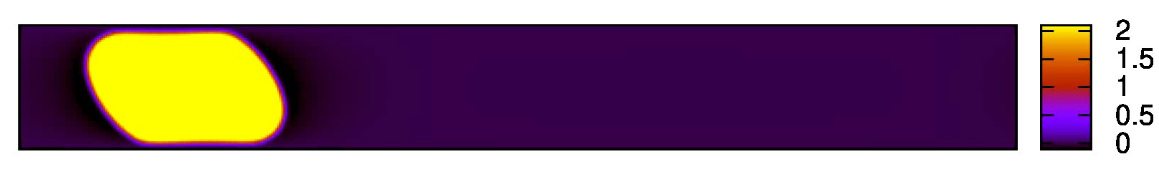}
\caption{Shapes and dynamic behavior of a polar active fluid droplet with $\lambda\simeq 0.7$ and $\zeta=10^{-3}$.  Here the drop is motionless. This state corresponds to the static regime 
of the phase diagram of Fig.10 (red/pluses) of the main text.}
\label{figS4}
\end{figure}

\bibliography{biblio}

\end{document}